\documentclass[twocolumn,showpacs,showkeys,superscriptaddress]{revtex4}
\usepackage{graphicx}
\usepackage{amsmath}
\usepackage{amsfonts}
\usepackage{amssymb}
\usepackage{mathrsfs}
\usepackage{color}

\def\openone{\leavevmode\hbox{\small1\kern-3.8pt\normalsize1}}
\def\N{\leavevmode\hbox{ Z \kern-8 pt\normalsize{Z}}}
\def\openone{\leavevmode\hbox{\small1\kern-3.8pt\normalsize1}}
\def\openJ{\leavevmode\hbox{J \kern-9.5pt\normalsize J}}
\def\openS{\leavevmode\hbox{ S \kern-9.3pt\normalsize S}}
\newcommand{\bb}{\begin{equation}}
\newcommand{\ee}{\end{equation}}
\newcommand{\eqb}{\begin{eqnarray}}
\newcommand{\eqf}{\end{eqnarray}}

\newcommand*{\field}[1]{\mathbb{#1}}%

\begin{document}

\title{Do electromagnetic waves always propagate along null geodesics?}

\author{Felipe A. Asenjo}
\email{felipe.asenjo@uai.cl}
\affiliation{Facultad de Ingenier\'{\i}a y Ciencias,
Universidad Adolfo Ib\'a\~nez, Santiago, Chile.}
\author{Sergio A. Hojman}
\email{sergio.hojman@uai.cl}
\affiliation{Departamento de Ciencias, Facultad de Artes Liberales,
Universidad Adolfo Ib\'a\~nez, Santiago, Chile.}
\affiliation{Departamento de F\'{\i}sica, Facultad de Ciencias, Universidad de Chile,
Santiago, Chile.}
\affiliation{Centro de Recursos Educativos Avanzados,
CREA, Santiago, Chile.}

\begin{abstract}
We find exact solutions to Maxwell equations written in terms of four-vector potentials in non--rotating, as well as in G\"odel and Kerr spacetimes. We show that Maxwell equations can be reduced to two uncoupled second-order differential equations for combinations of the components of  the  four-vector potential.
Exact electromagnetic waves solutions are written on given gravitational field backgrounds where they evolve. We find that in non--rotating spherical symmetric spacetimes, electromagnetic waves travel along null geodesics. However, electromagnetic  waves on G\"odel and Kerr spacetimes do not exhibit that behavior.
\end{abstract}


\maketitle

\section{Introduction}

The Equivalence Principle (EP) is one of the cornerstones of General Relativity. The EP is usually invoked when the propagation of light is studied.  It is important to state that the EP applies to point--like objects only, i.e., to objects which are completely described by their spacetime position. Any particle with structure experiences tidal gravitational forces which modify its trajectories. Therefore, one can wonder if the EP can be used to  fully explain the motion of extended particles, such as particles with spin \cite{gane1,gane2,nicolas,hanson,hojmantesis}. Similarly, waves are extended physical objects, and in addition, photons have spin, which is related to the fact that electromagnetic (EM) waves have different polarization states which, in general, modify their propagation. Therefore, when the characteristic scales of the gravitational field and the wave are comparable,
the EP cannot be used to fully understand wave dynamics. The applicability of the EP on the study of EM waves is valid only in the high-frequency limit, where EM radiation is better described by a massless particle which travels along null geodesics than by a wave model.

To unveil the realm of validity of the EP, we study the propagation of electromagnetic waves on curved spacetimes. Maxwell equations are
\begin{equation}\label{eMax0a}
\nabla_\alpha F^{\alpha\beta}=0\, ,\qquad \nabla_\alpha F^{*\alpha\beta}=0\, ,
\end{equation}
 where $\nabla_\alpha$ stands for the covariant derivative defined for a metric $g_{\mu\nu}$,
and  $F^{\alpha\beta}$ is the antisymmetric electromagnetic field tensor, while $F^{*\alpha\beta}$ is its dual. If the electromagnetic field is written in terms of the four-vector potentials $A_\alpha$, i.e., $F^{\alpha\beta}=g^{\alpha \mu} g^{\beta \nu}F_{\mu \nu}$ and
$F_{\mu\nu}=\nabla_\mu A_\nu-\nabla_\nu A_\mu$, then $F_{\mu\nu}=\partial_\mu A_\nu-\partial_\nu A_\mu$, where $\partial_\mu$ is a partial derivative. With these definitions, the equations  $\nabla_\alpha F^{*\alpha\beta}=0$ are identically satisfied. The Eqs.~\eqref{eMax0a} to be solved are
\begin{equation}\label{eMax1}
\partial_\alpha\left[\sqrt{-g}g^{\alpha\mu}g^{\beta\nu} (\partial_\mu A_\nu-\partial_\nu A_\mu)\right]=0\, ,
\end{equation}
where $g^{\mu\nu}$ is the inverse metric.

We deal with test electromagnetic fields which evolve on a given gravitational background field. Several exact solutions for Maxwell equations in curved spacetimes have been found \cite{plebanski,teu,chandra,felice,mashhoon,mashhoon2,cohen,tsagas}. One of the most interesting solutions of Eq.~\eqref{eMax1},  due to their physical relevance, are electromagnetic  waves.
We take  waves to be described by $A_\mu=\xi_\mu e^{i{S}}$   (or its real part), where $\xi_\mu$ is the amplitude and ${ S}$ the phase of the wave   \cite{misner,born}. Both are real quantities and, in principle, both depend on space and time. The four-wavevector of the wave is defined by
$K_\mu=\nabla_\mu { S}=\partial_\mu { S}$,
where $K_0$ is identified with the frequency of the wave, whereas $K_i$ are the components of the (three dimensional) wavevector.
The nature of the propagation of the electromagnetic wave is determined by a constraint satisfied by the four dimensional wave. For example, in vacuum flat-spacetime $K_\mu K^\mu=0$, and electromagnetic radiation evolves along null geodesics, i.e., the wave travels with the speed of light.

Usually, Maxwell equations \eqref{eMax1} are solved using the geometrical optics approximation (high-frequency limit) \cite{plebanski,felice,misner,born,tolman}, where the wavelength of the wave is considered much smaller than any characteristic length scale of the gravitational field  in which the wave evolves. Using this approximation, Maxwell equations \eqref{eMax1} are  solved perturbatively, and in this approximation, the electromagnetic plane wave solutions are described by \cite{misner}
\begin{equation}\label{solGeom}
K_\mu K^\mu=0\, ,\quad \xi_\mu K^\mu=0\, ,\quad \nabla_\mu\left(K^\mu \xi^\alpha \xi_\alpha\right)=0\, .
\end{equation}
The first equation implies that plane waves follow null geodesics, whereas the second one shows that the wave is transverse. The third one is the photon number conservation equation.
Despite the simplicity of solution \eqref{solGeom}, there is no guarantee that an exact solution to Eqs.~\eqref{eMax1} will satisfy the same conditions \eqref{solGeom} for any given gravitational field.

The purpose is this work is to show that there exist particular wave solutions to Maxwell equations which do not evolve along null geodesics. Wave solutions have been studied previously \cite{teu,chandra,felice,mashhoon,mashhoon2}.  Nevertheless, these results obtained previously do not include solutions such as the ones we present here, i.e., $A_\mu=\xi_\mu e^{i{S}}$ light waves which may propagate along non--null trajectories due to rotation--polarization interaction, which as far as we know, is the first time that have been reported in the literature (in conjunction with a companion article \cite{bianchiasenjohojman}, where the non--null trajectory effect is due to anisotropy--polarization interaction.) In several of these works  \cite{felice,mashhoon,mashhoon2}, the analogy of the gravitational field with a medium (with its corresponding susceptibility and permeability) is exhibited in an explicit way.  As electromagnetic waves in a physical medium can travel at speeds different from the speed of light in vacuum, and also present effects such as  dispersion or birefringence, among others, one can wonder if the gravitational field can also modify the nature of propagation of light on curved spacetimes. Therefore, it is the purpose of this work to find out whether electromagnetic waves  can propagate along paths which are not null geodesics, and if other effects can emerge from the coupling between electromagnetism and gravity.

\section{Electromagnetic waves in non--rotating spherically symmetric spacetimes}

 Before proceeding to find solutions propagating in G\"odel and Kerr metrics, we show how the null geodesics propagation of light emerges from Maxwell equations \eqref{eMax1} in non--rotating spherically symmetric spacetimes.

The following solutions are exact (i.e., they are obtained without using the eikonal approximation or any other kind of approximation) and they show that there are EM waves that travel at the speed of light, independent of their scales, for some spacetime backgrounds.

 Examples of these background gravitational fields are Schwarzschild, Reissner-Nordstrom \cite{misner}, Friedmann--Robertson--Walker (FRW) \cite{ryden} and wormholes \cite{visser}, for instance.  We consider a general symmetric metric in spherical coordinates $(t=x^0,r,\theta,\phi)$, such that
\begin{eqnarray}
g_{00}&=&f(t)q(r)\, ,\nonumber\\
g_{rr}&=&h(t)b(r)\, ,\nonumber\\
g_{\theta\theta}&=&h(t)r^2\, ,\nonumber\\
g_{\phi\phi}&=&h(t)r^2\sin^2\theta\, ,
\end{eqnarray}
 where $f(t), h(t),  q(r)$ and $b(r)$ are (up to now) arbitrary functions of $t$ and $r$ respectively. All other metric components vanish.

We can solve Eqs.~\eqref{eMax1} in the Weyl gauge $A_0=0$ for simplicity (although they can be solved without imposing any gauge).
The time-component of Eqs.~\eqref{eMax1} becomes a vanishing curl statement and thus it can be identically solved by introducing a new field $\chi$ that satisfies
\begin{eqnarray}\label{chiScha1}
\partial_\theta \chi &=&-\sqrt{-\frac{h}{fqb}}r^2\sin\theta\, \partial_0 A_r\, ,\nonumber\\
\partial_r \chi &=&\sqrt{-\frac{hb}{fq}}\sin\theta\, \partial_0 A_\theta\, ,
\end{eqnarray}
since we have chosen to deal with fields with no $\phi$-dependence of the fields (the $\phi$-dependence can be determined by studying fields which are proportional to $e^{im\phi}$ with an arbitrary parameter $m$).

 Similarly, the $r$- and $\theta$-components of   Eqs.~\eqref{eMax1} are identically satisfied by the relation
\begin{equation}\label{chiScha2}
\sqrt{-\frac{h}{f}}\, \partial_0\chi=\sqrt{\frac{q}{b}}\sin\theta\left(\partial_r A_\theta-\partial_\theta A_r\right)\, .
\end{equation}
Eqs.~\eqref{chiScha1} and \eqref{chiScha2} can be combined to produce the wave equation for the $\chi$ field
\begin{eqnarray}\label{eqSchwchi}
&&-\sqrt{-\frac{h}{f}}\partial_0\left(\sqrt{-\frac{h}{f}}\partial_0 \chi\right)+\sqrt{\frac{q}{b}}\partial_r\left(\sqrt{\frac{q}{b}}\partial_r \chi\right)\nonumber\\
&&\qquad\qquad\qquad\qquad +\frac{q}{r^2}\sin\theta\, \partial_\theta\left(\frac{\partial_\theta\chi}{\sin\theta}\right)=0\, .
\end{eqnarray}

On the other hand, the $\phi$-component of Eqs.~\eqref{eMax1} decouples
\begin{eqnarray}\label{eqSchw}
&&-\sqrt{-\frac{h}{f}}\partial_0\left(\sqrt{-\frac{h}{f}}\partial_0 A_\phi\right)+\sqrt{\frac{q}{b}}\partial_r\left(\sqrt{\frac{q}{b}}\partial_r A_\phi\right)\nonumber\\
&&\qquad\qquad\qquad\qquad +\frac{q}{r^2}\sin\theta\, \partial_\theta\left(\frac{\partial_\theta A_\phi}{\sin\theta}\right)=0\, .
\end{eqnarray}

Eqs.~\eqref{eqSchwchi} and \eqref{eqSchw} have the same form. In general, defining new time and radial coordinates as $\tau =\int dt \sqrt{-f/h}$, and $\rho=\int dr \sqrt{b/q}$, the previous equations become the flat-spacetime wave equation $\partial_\tau^2 A_\phi-\partial_\rho^2 A_\phi=0$, where the $A_\phi$ solution may be written in a  wave fashion as
\begin{equation}
A_\phi=\cos\theta\exp\left(i \omega \int dt \sqrt{-\frac{f}{h}}\pm i \omega\int dr \sqrt{\frac{b}{q}}\right)\, ,
\end{equation}
where $\omega$ is a constant. The field $\chi$ has the same wave solution.
This solution defines the wave-vectors  $K_0=\omega\sqrt{-{f}/{h}}$, and $K_r=\pm\omega\sqrt{{b}/{q}}$,
which satisfy
\begin{equation}\label{soluSch}
K_\mu K^\mu=g^{00}K_0^2+g^{rr}K_r^2=\frac{K_0^2}{f q}+\frac{K_r^2}{hb}=0\, .
\end{equation}
Therefore, all non--rotating spherically symmetric spacetimes have electromagnetic  wave solutions travelling along null geodesics which are transversal waves $A_\mu K^\mu=0$.
In general for this case, if  $A_\phi(t,r, \theta)=\cos\theta \xi(r) e^{i {S}(t,r)}$ is used in Eq.~\eqref{eqSchw}, where $K_0=\partial_t {S}$ and $K_r=\partial_r S$, then  for constant amplitude $\xi$, the wave  follows null geodesics as in \eqref{soluSch} and the photon number is conserved \cite{misner}.

Let us consider the Schwarzschild and FRW spacetimes. The null-geodesic behavior of the light \eqref{soluSch}, implies that for Schwarzschild spacetime the  wave dispersion relation is
\begin{equation}
K_0=\pm\left(1-\frac{2M}{r}\right)K_r\, .
\end{equation}
This solution corresponds to the well-known gravitational redshift effect. This  dispersion relation can  be written in the wave time $\tau$ as
\begin{equation}
\omega=\frac{\partial S}{\partial \tau}=\frac{\partial t}{\partial \tau}K_0=\frac{K_0}{1-2M/r}=\pm K_r\, ,
\end{equation}
from where it can be obtained that the wave moves with group velocity $\partial\omega/\partial K_r=\pm 1$ at the speed of light.

Similarly, for  the FRW spacetime, the wave disperses as
\begin{equation}
K_0=\pm\frac{\sqrt{1-k r^2}}{a}K_r\, ,
\end{equation}
which is the cosmological redshift for a Universe with curvature $k=-1,0,1$.  In wave time $\tau$, the wave has the dispersion relation
\begin{equation}
\omega=\frac{\partial S}{\partial \tau}=\frac{a K_0}{\sqrt{1-k r^2}}=\pm K_r\, ,
\end{equation}
and in the FRW Universe the wave propagates at the speed of light $\partial\omega/\partial K_r=\pm 1$.

 \section{Electromagnetic waves in G\"odel spacetime}

 The G\"odel metric describes a rotating Universe
which features closed timelike curves. This metric is  stationary \cite{mashhoon2,hawk}  and its components can be written in cartesian coordinates as
\begin{eqnarray}
g_{00}&=&-1=-g_{xx}=-g_{zz}\, ,\nonumber\\
g_{yy}&=&-2+4 \exp(\sqrt{2} x\Omega)-\exp(2\sqrt{2}x\Omega)\, ,\nonumber\\
g_{0y}&=&\sqrt{2}[1-\exp(\sqrt{2} x \Omega)]\, ,
\end{eqnarray}
where $\Omega$ is a constant related to the angular velocity of the rotating
universe (which reaches the flat spacetime limit when $\Omega\rightarrow 0$). All other metric components vanish.

The propagation of electromagnetic  waves can be studied in a general  fashion similar to the previous section. However, to explicitly show a simple solution for EM wave, we restrict ourselves to a particular solution of Eq.~\eqref{eMax1}. Choosing $A_\mu(t,x)=A_z(t,x) \delta_{\mu z}$, Maxwell equations reduce to
\begin{equation}\label{Go1}
\partial_0^2 A_z+\frac{1}{\sqrt{-g}g^{00}}\partial_x\left(\sqrt{-g}\partial_x A_z\right)=0\, ,
\end{equation}
where $g=-\exp(2\sqrt{2}x\Omega)$ is the metric determinant and $g^{00}=1+2 \exp(-2\sqrt{2} x\Omega)-4\exp(-\sqrt{2}x\Omega)$.
One can  define the variable $\zeta=-e^{-\sqrt{2} x\Omega}/(\sqrt{2}\Omega)$, to rewrite the Eq.~\eqref{Go1} as $\partial_0^2 A_z+\sigma(\zeta) \partial^2_\zeta A_z=0$, where $\sigma(\zeta)={2\Omega^2\zeta^2}/({1+4\Omega^2\zeta^2+4\sqrt{2}\Omega\zeta})$.
We can see that the wave equation cannot be cast in a flat spacetime analogue version, and thereby, the null geodesic behavior of the light in G\"odel spacetimes is ruled out.  The function $\sigma$ emerges as an effective wave velocity due to the spacetime rotation.
To explicitly show this, let us go back to Eq.~\eqref{Go1} and perform the  wave ansatz $A_z(t,x)=\xi(x) \exp[i\omega t \pm i S(x)]$, where $\xi$ is the wave amplitude and $\omega$ is a constant. The wavevectors are derivatives of the phase, $K_0=\omega$ and $K_x=\pm\partial_x S$, which allow us to describe a transversal electromagnetic  wave $A_\mu K^\mu\equiv 0$.
Using this ansatz, Eq.~\eqref{Go1} becomes
\begin{eqnarray}\label{setGodel}
K_\mu K^\mu&=&\frac{1}{\xi\sqrt{-g}}\partial_x\left(\sqrt{-g} \partial_x \xi\right)\, ,\nonumber\\
0&=&\partial_x\left(\sqrt{-g}K_x \xi^2\right)\, ,
\end{eqnarray}
where $K_\mu K^\mu\equiv g^{00} \omega^2+K_x^2$. The first equation is the dispersion relation of the wave determining the  behavior of light. The second one is the   photon number conservation,
and it can be solved exactly to get
\begin{equation}\label{solxigo}
\xi=\frac{\xi_0}{(-g)^{1/4}K_x^{1/2}}\, ,
\end{equation}
with  arbitrary $\xi_0$.

First, one can notice that in the high--frequency limit the amplitude variations are neglected compared to wave scale lengths \cite{misner}. Then, the dispersion relation \eqref{Go1} becomes $K_\mu K^\mu=0$.

However, if one tries to solve the system exactly, it can be shown that there are no consistent exact solutions of the previous system with constant amplitude or with $K_\mu K^\mu=0$. For the latter case, Eqs.~\eqref{setGodel} become three different and inconsistent conditions for the two variables $K_x$ and $\xi$.

One can use solution \eqref{solxigo} in the first of Eqs.~\eqref{setGodel} to get the dispersion relation
\begin{eqnarray}\label{setGodel2}
g^{00} \omega^2+K_x^2=-\frac{\Omega^2}{2}
-\frac{K_x''}{2K_x}+\frac{3K_x'^2}{4 K_x^2} \, ,
\end{eqnarray}
where $'\equiv \partial_x$. From the previous exact dispersion relation, we can see that for G\"odel  spacetime this particular electromagnetic  wave does not propagate in null geodesics. The exact geodesic behavior can be found by solving the differential equation \eqref{setGodel2} for $K_x$.  This feature has its origin in the rotation of the spacetime, which modifies the path followed by photons. The null geodesic flat spacetime behavior can be obtained when $\Omega$ vanishes, as $g^{00}\rightarrow -1$, $g\rightarrow -1$, and $K_x\rightarrow\omega$ constant.

It is illustrative to calculate the solutions of \eqref{setGodel2}  in the case $\Omega \, x\ll 1$. At second order in $\Omega$, the solution to \eqref{setGodel2} is
\begin{equation}\label{solGodel1}
K_x\approx\omega-2\Omega^2 x^2 \omega+\frac{3\Omega^2}{2\omega}\sin^2(\omega x)\, ,
\end{equation}
that allow us to recover the flat spacetime solution when $x=0$ (the metric becomes flat at that point). This solution implies that
\begin{equation}
K_\mu K^\mu=3\Omega^2\sin^2(\omega x)\, ,
\end{equation}
 which is always positive. Thus, electromagnetic  waves in a slowly rotating G\"odel Universe propagates in space-like trajectories. This can be easily seen at small spacetime length scales $\omega x\ll1$, such that solution \eqref{solGodel1} simplifies to
\begin{equation}\label{solGodel1approx}
{\omega}\approx \left(1+\frac{1}{2}\Omega^2 x^2\right){K_x}\, ,
\end{equation}
and the  waves propagate with  superluminal  group velocity
\begin{equation}
\frac{\partial \omega}{\partial K_x}\approx 1+\frac{1}{2}\Omega^2 x^2\geq 1\, ,
\end{equation}
for $0\leq \Omega x\ll 1$. This  wave moves at the speed of light in the flat spacetime limit $\Omega\rightarrow 0$.

This somewhat surprising behavior of an electromagnetic  wave solution stems from the rotational character of the spacetime under consideration. This phenomenon may be closely related to the fact that this metric admits closed timelike curves.
Electromagnetic waves in G\"odel spacetimes have been studied by Mashoon \cite{mashhoon2} in a general formalism. However, no explicit  wave solutions were  presented.

The EM field invariants can be readily obtained for the G\"odel case. The approximated solution \eqref{solGodel1approx}, with $\Omega x, \omega x\ll1$,  gives rise to  an electric field $F_{0z}=\partial_0 A_z$, and a magnetic field $F_{xz}=\partial_x A_z$, both describing a non-null electromagnetic field in general. With the above solution, the invariant $F^{*\mu\nu}F_{\mu\nu}=0$, representing a transverse wave. However, the other invariant
$F^{\mu\nu}F_{\mu\nu}\propto \omega\Omega\sin[2\omega(t+x)]$  does not vanish in general. For $t+x=n\pi/(2\omega)$, with $n\in \field{N}$, the EM field is null. On the contrary, for $0<t+x< \pi/(2\omega)$ the EM field has a region of magnetic dominance, and for $\pi/(2\omega)<t+x< \pi/\omega$ the region is of electric dominance.
 In the flat spacetime limit, the EM field becomes always a null field.

Because of these new results regarding wave propagation in G\"odel spacetimes, it is natural to inquire whether propagation of electromagnetic  waves in Kerr spacetime presents similar features.

\section{Electromagnetic waves in Kerr spacetime}

The stationary Kerr metric describes a rotating black hole of mass $M$ and effective angular momentum $a$. It has non-vanishing metric components $g_{\mu\mu}=g_{\mu\mu}(r,\theta)$ and $g_{\phi 0}=g_{0\phi}=g_{0\phi}(r,\theta)$,  with $\mu=\{t=x^0, r, \theta, \phi\}$.
Explicitly, the metric in Boyer-Lindquist coordinates is
\begin{eqnarray}
g_{00}&=&-1+2M r/\rho^2\, ,\nonumber\\
g_{rr}&=& \rho^2/\Delta\, ,\nonumber\\
g_{\theta\theta}&=&\rho^2\, ,\nonumber\\
g_{\phi\phi}&=&r^2+a^2+2 M r a^2\sin^2\theta/\rho^2\, ,\nonumber\\
g_{0\phi}&=&-4 M a  r\sin^2\theta/\rho^2\, ,
\end{eqnarray}
where
$\rho^2=r^2+a^2\cos^2\theta$ and $\Delta=r^2-2Mr+a^2$.
In contrast to the two previous cases, now the metric depends on two spatial variables. We now show below that Maxwell equations \eqref{eMax1} can be solved exactly for all of the potential components (without choosing a gauge) for the Kerr metric.

Write the four-vector potential components for this case as $A_\mu(t, r,\theta)$, with no $\phi$-dependence, for simplicity. With this assumption, the time-component of Maxwell equation \eqref{eMax1} can be understood as a vanishing curl statement and may thus be solved by introducing a new field $\chi=\chi(t, r,\theta)$ such that
\begin{eqnarray}\label{chi1}
\partial_\theta\chi&=&-\sqrt{-g}g^{rr}\left[g^{00}\left(\partial_rA_0-\partial_0 A_r\right)+g^{0\phi}\partial_r A_\phi\right]\, ,\nonumber\\
\partial_r\chi&=&\sqrt{-g}g^{\theta\theta}\left[g^{00}\left(\partial_\theta A_0-\partial_0 A_\theta\right)+g^{0\phi}\partial_\theta A_\phi\right]\, ,
\end{eqnarray}
which identically satisfies the time-component of Eq.~\eqref{eMax1}.
The introduction of the new field $\chi$  also allows us to solve the $r$ and $\theta$-components of Eq.~\eqref{eMax1}. Both equations reduce to one equation, namely,
\begin{equation}\label{chi2}
\partial_0\chi=\sqrt{-g}g^{rr}g^{\theta\theta}\left(\partial_r A_\theta -\partial_\theta A_r\right)\, .
\end{equation}
Finally, the $\phi$-component of Maxwell equations may be written as
\begin{eqnarray}\label{phiMax}
0&=&\sqrt{-g}\frac{g^{00}}{g_{\phi\phi}}\partial_0^2 A_\phi+\partial_r\left(\sqrt{-g}\frac{g^{rr}}{g_{\phi\phi}}\partial_r A_\phi\right)\nonumber\\
&&+\partial_\theta\left(\sqrt{-g}\frac{g^{\theta\theta}}{g_{\phi\phi}}\partial_\theta A_\phi\right)-\partial_r\beta\partial_\theta\chi+\partial_\theta\beta\partial_r\chi\, ,
\end{eqnarray}
where $\beta=g^{0\phi}/g^{00}=-g_{0\phi}/g_{\phi\phi}$ is related to the rotation rate of the black hole. On the other hand, as the metric is time-independent, from Eqs.~\eqref{chi1} and \eqref{chi2} we can find an evolution equation for the $\chi$ field
\begin{eqnarray}\label{chiMax}
0&=&\sqrt{-g}\frac{g^{00}}{g_{\phi\phi}}\partial_0^2 \chi+\partial_r\left(\sqrt{-g}\frac{g^{rr}}{g_{\phi\phi}}\partial_r \chi\right)\nonumber\\
&&+\partial_\theta\left(\sqrt{-g}\frac{g^{\theta\theta}}{g_{\phi\phi}}\partial_\theta \chi\right)+\partial_r\beta\partial_\theta A_\phi-\partial_\theta\beta\partial_r A_\phi\, .
\end{eqnarray}
Now, definining the complex potential $Z_{\pm}=A_\phi \pm i\chi$, Eqs.~\eqref{phiMax}  and \eqref{chiMax} can be merged to
\begin{eqnarray}\label{ZMax}
0&=&\sqrt{-g}\frac{g^{00}}{g_{\phi\phi}}\partial_0^2 Z_\pm+\partial_r\left(\sqrt{-g}\frac{g^{rr}}{g_{\phi\phi}}\partial_r Z_\pm\right)\nonumber\\
&&+\partial_\theta\left(\sqrt{-g}\frac{g^{\theta\theta}}{g_{\phi\phi}}\partial_\theta Z_\pm\right)\pm i\partial_r\beta\partial_\theta Z_\pm \mp i \partial_\theta\beta\partial_r Z_\pm\, .\nonumber\\
&&
\end{eqnarray}

In this way, the problem of getting the solutions to the four Maxwell equations \eqref{eMax1}, is now reduced to solve the two uncoupled equations  \eqref{ZMax}. Different from the cases on Sec.~II, now it is the variations of spacetime rotation that couples the components of the EM potential.
The different space and time derivatives of the fields $Z_\pm$ can be related to the polarizations of the EM wave. Thus, Eqs.~\eqref{ZMax} take into account the polarization of the  wave. This implies that different EM polarizations can couple to the black hole rotation through the derivatives of $\beta$.
If the spacetime is static $\beta=0$ (as in Schwarzschild case) this effect does not appear. This feature has been previously envisaged \cite{mashhoon2}, but no exact solution for a  wave was presented there.


Now, to find an EM wave solution, the polarization function is written as
$Z_\pm(t,r,\theta)=\xi_\pm(r,\theta) e^{i\omega t + i S_\pm(r,\theta)}$, where $\omega$ is a constant and  $\xi_\pm$ is the amplitude of the wave (in inverse length units). Anew, the four wavevector components are defined as ${K_0}_\pm=\omega$, ${K_r}_\pm=\partial_r S_\pm$, ${K_\theta}_\pm=\partial_\theta S_\pm$, and ${K_\phi}_\pm=0$.
Also notice that
\begin{equation}\label{notransversAK}
A_\mu {K^\mu}_\pm=g^{00}\omega A_0+g^{rr}{K_r}_\pm A_r+g^{\theta\theta}{K_\theta}_\pm A_\theta\neq0\, ,
\end{equation}
in general for Kerr spacetime and, therefore, the wave is not transverse.
Using this decomposition, Eq.~\eqref{ZMax}  gives rise to the dispersion relation
\begin{eqnarray}\label{DispK}
&&\frac{\sqrt{-g}}{g_{\phi\phi}}{K^\mu}_\pm {K_\mu}_\pm=\mp\partial_r\beta {K_\theta}_\pm \pm \partial_\theta\beta {K_r}_\pm\nonumber\\
&&+\frac{1}{\xi_\pm}\partial_r\left(\sqrt{-g}\frac{g^{rr}}{g_{\phi\phi}}\partial_r\xi_\pm\right)+\frac{1}{\xi_\pm}\partial_\theta\left(\sqrt{-g}\frac{g^{\theta\theta}}{g_{\phi\phi}}\partial_\theta\xi_\pm\right)\, ,\nonumber\\
&&\end{eqnarray}
where ${K^\mu}_\pm {K_\mu}_\pm\equiv g^{00}\omega^2+g^{rr}{K_r}^2_\pm+g^{\theta\theta}{K_\theta}^2_\pm$. Also it gives origin to the generalized photon number conservation law for Kerr spacetime given by
\begin{eqnarray}\label{ConsK1}
0&=&\partial_r\left(\sqrt{-g}\frac{g^{rr}}{g_{\phi\phi}}{K_r}_\pm\right)+\partial_\theta\left(\sqrt{-g}\frac{g^{\theta\theta}}{g_{\phi\phi}}{K_\theta}_\pm\right)\nonumber\\
&&+\left(2\sqrt{-g}\frac{g^{rr}}{g_{\phi\phi}}{K_r}_\pm\mp\partial_\theta\beta\right)\frac{\partial_r\xi_\pm}{\xi_\pm}\nonumber\\
&&+\left(2\sqrt{-g}\frac{g^{\theta\theta}}{g_{\phi\phi}}{K_\theta}_\pm\pm\partial_r\beta\right)\frac{\partial_\theta\xi_\pm}{\xi_\pm}\, ,
\end{eqnarray}
which can be cast in a more appealing fashion
\begin{eqnarray}\label{ConsK2}
0&=&\partial_r\left[\left(\sqrt{-g}\frac{g^{rr}}{g_{\phi\phi}}{K_r}_\pm\mp\frac{1}{2}\partial_\theta\beta\right){\xi^2_\pm}\right]\nonumber\\
&&+\partial_\theta\left[\left(\sqrt{-g}\frac{g^{\theta\theta}}{g_{\phi\phi}}{K_\theta}_\pm\pm\frac{1}{2}\partial_r\beta\right){\xi^2_\pm}\right]\, .
\end{eqnarray}

Two important limits must be analyzed. One must notice that as $\beta\rightarrow 0$, Eqs.~\eqref{ConsK1} and \eqref{ConsK2} reduce to the Schwarzschild case discussed in Sec.~II, where $\xi\rightarrow\cos\theta$ and $K_\theta\rightarrow 0$, and the EM propagates in null geodesics. On the other hand, on the high-frequency limit, the scale lengths of the wave amplitude and the spacetime are much larger than those of the EM wave, and the gradient of $\beta$ and $\xi$ are neglected in \eqref{DispK}.  This is the eikonal limit, and the EM wave travels on null geodesics ${K^\mu}_\pm {K_\mu}_\pm=0$, whereas the dynamics of the amplitude is determined by Eqs.~\eqref{ConsK1} or \eqref{ConsK2}.

Nonetheless, we can show that Eq.~\eqref{ConsK2} can be solved exactly and use it, in principle, to determine the value of $K^\mu K_\mu$ by the dynamical equations. A consistent solution of Eq.~\eqref{ConsK1} [or \eqref{ConsK2}] should specify the behavior of a  wave travelling on a Kerr background.
 A particular solution to  Eq.~\eqref{ConsK2} is
\begin{eqnarray}\label{setxi}
\sqrt{-g}\frac{g^{rr}}{g_{\phi\phi}}{K_r}_\pm\mp\frac{1}{2}\partial_\theta\beta&=&\partial_\theta \xi_\pm+\frac{\lambda(\theta)}{\xi^2_\pm}\, ,\nonumber\\
\sqrt{-g}\frac{g^{\theta\theta}}{g_{\phi\phi}}{K_\theta}_\pm\pm\frac{1}{2}\partial_r\beta&=&-\partial_r \xi_\pm\, ,
\end{eqnarray}
where $\lambda$ is an arbitrary function of $\theta$. This solution has the correct  Schwarzschild spacetime limit under the choices of  $\xi=\xi_0$, $\beta\rightarrow 0$, $K_\theta\rightarrow 0$ and $\lambda=\omega\xi_0^2/\sin\theta$. From Eqs.~\eqref{setxi} we are able to find a solution for the wave amplitude that  can be obtained by manipulating the equations of that set. In general, from Eqs.~\eqref{setxi}, we get (for a non-constant amplitude)
\begin{widetext}
\begin{eqnarray}\label{solxi}
\xi_\pm&=&\lambda^{1/3}\left[2\sqrt{-g}\frac{g^{\theta\theta}}{g_{\phi\phi}}{K_\theta}_\pm\pm\partial_r\beta \right]^{1/3}\left[ \partial_r\left(\sqrt{-g}\frac{g^{rr}}{g_{\phi\phi}}{K_r}_\pm\right)+\partial_\theta\left(\sqrt{-g}\frac{g^{\theta\theta}}{g_{\phi\phi}}{K_\theta}_\pm\right) \right]^{-1/3}\, .
\end{eqnarray}
It is remarkable that this amplitude solves the photon number conservation equation in Kerr spacetime. The same result may be gotten using Eq.~\eqref{ConsK1}.
Thereby, solutions \eqref{setxi} and \eqref{solxi} can be used in dispersion relation \eqref{DispK} to finally yield
\begin{eqnarray}\label{DispK2}
\frac{\sqrt{-g}}{g_{\phi\phi}}{K^\mu}_\pm {K_\mu}_\pm&=&\mp\partial_r\beta {K_\theta}_\pm \pm \partial_\theta\beta {K_r}_\pm
+\frac{1}{\xi_\pm}\left[ \partial_\theta\left(\beta^2-\frac{g_{00}}{g_{\phi\phi}}\right) K_{r\pm} -\partial_r\left(\beta^2-\frac{g_{00}}{g_{\phi\phi}}\right) K_{\theta\pm}\right]\nonumber\\
&&\mp\frac{1}{\xi_\pm}\left[\partial_r\left(\frac{\sqrt{-g} g^{rr}}{2g_{\phi\phi}}\partial_r\beta\right) + \partial_\theta\left(\frac{\sqrt{-g} g^{\theta\theta}}{2g_{\phi\phi}}\partial_\theta\beta\right)   \right]-\frac{1}{\xi_\pm^3}\partial_\theta \left(\frac{\sqrt{-g} g^{\theta\theta}\lambda}{g_{\phi\phi}}\right)\mp \frac{\sqrt{-g} g^{\theta\theta}\lambda\partial_\theta\beta}{g_{\phi\phi}\xi_\pm^4}\nonumber\\
&&+\frac{2\lambda}{ \xi_\pm^4}\left(\beta^2-\frac{g_{00}}{g_{\phi\phi}}\right)K_{r\pm}-\frac{2\sqrt{-g} g^{\theta\theta} \lambda^2}{g_{\phi\phi}\xi_\pm^6}\, ,
\end{eqnarray}
\end{widetext}
where the terms $\xi_\pm$, $\xi_\pm^3$, $\xi_\pm^4$ and $\xi_\pm^6$ in \eqref{DispK2} must be replaced using Eq.~\eqref{solxi}.

We can see from the above dispersion relation that the  wave has ${K^\mu}_\pm {K_\mu}_\pm\neq 0$, and consequently the electromagnetic  wave does not travel along  null geodesics and exhibits birefringence. The proper behavior of the wave can be found by solving the differential equation \eqref{DispK2} for $S_\pm$. It is worth noting that both evolution equations for $Z_{+}$ and $Z_{-}$ couple differently to the derivatives of $\beta$ so, the solutions for each polarization state are different, in general. This effect is a direct consequence of the coupling of light polarization and the rotation of the central mass. In this way, light can travel along different paths at different speeds depending on its polarization. This an effect suggested by Mashhoon \cite{mashhoon2} and it is analogous to the well known Faraday rotation effect in plasmas \cite{chen}. On the other hand, this polarization-gravity coupling effect is
intimately related with photon spin coupling to gravitational fields  studied in Ref.~\cite{nicolas} from a classical viewpoint.

An approximate solution can be found for slowly rotating $a\ll 1$ black holes  with EM waves propagating in such a way that wave vectors are almost parallel to the black hole angular momentum vector. For an EM wave with constant frequency  $\omega$, the solution of Eqs.~\eqref{DispK} and \eqref{ConsK1},
for $\theta\approx 0$, has an amplitude with the form
\begin{equation}\label{approximatedKerr1}
\xi_\pm\approx \cos\theta \left[1\mp\frac{a\, \eta}{2\omega^3}\cos\left(2\omega\left[r+2M\ln(r-2M)\right]\right)\right]\, ,
\end{equation}
 (where $\eta$ is a constant) and wave-vectors
\begin{eqnarray}
K_{r\pm}&\approx& \frac{r\omega}{r-2M}\left[1\pm\frac{a\, \eta}{\omega^3}\cos\left(2\omega\left[r+2M\ln(r-2M)\right]\right)\right]\, ,\nonumber\\
K_{\theta\pm}&\approx& 0\, .
\end{eqnarray}

Notice that the solutions are different because of the polarization of the EM wave.
This  implies that  the EM wave propagation is timelike or spacelike depending on the polarization
\begin{equation}\label{approximatedKerr3}
{K^\mu}_{\pm} K_{\mu\pm}\approx \pm \frac{2 r a\, \eta}{\omega(r-2M)}\cos\left(2\omega\left[r+2M\ln(r-2M)\right]\right)\, .
\end{equation}
For $a=0$, we recover the null geodesic behavior of the propagation of EM waves on a Schwarzschild background. On the other hand, in the eikonal limit $\omega\rightarrow\infty$  we also recover the null geodesic behavior ${K^\mu}_{\pm} K_{\mu\pm}=0$.

Finally, the above (approximate) wave solution represents a non-null electromagnetic field in general, as the two EM field invariants are now proportional to $a$. With the assumptions used to obtain the solutions \eqref{approximatedKerr1}--\eqref{approximatedKerr3}, we obtain the invariant $F^{*\mu\nu}F_{\mu\nu}\propto a {\cal F}(t,r)$, where ${\cal F}=r \sin(\omega {\cal Y})[\cos(\omega(3{\cal Y}-2t))+\sin(\omega(3{\cal Y}-2t))]/(\omega(r-2M))$ is a function on time, radial distance to the black hole, and ${\cal Y}\equiv r+t+2M\ln(r-2M)$. Similarly, the invariant
$F^{\mu\nu}F_{\mu\nu}\propto  a {\cal G}(t,r)$, where ${\cal G}=\cos(2\omega({\cal Y}-t))\sin(2\omega{\cal Y})/(r\omega^2(r-2M))$. In general, both invariants do not vanish. Under the above approximations, this non-null field implies that in some reference frame, the electric and magnetic fields can be parallel. This  is related to the non-transversality of the wave \eqref{notransversAK}. However, the particular behavior of the invariants strongly depend on the functions ${\cal F}$ and ${\cal G}$. For instance, when ${\cal Y}=n\pi/\omega$, with $n\in \field{N}$, both ${\cal F}=0={\cal G}$, and the EM wave becomes a null field. Otherwise, the  EM field can present regions of electric or magnetic dominance. Lastly, in the Schwarzschild spacetime limit, the EM field becomes a null field always.

\section{Discussion}

DeWitt and Brehme \cite{dwb} obtained in 1960 results which are similar, but not equivalent, to ours. They used bitensors to properly define Green's functions to solve the scalar and Maxwell equations on a given general curved background. Their solutions exhibit ``tails'' inside the light cone which were attributed to scattering of the scalar and Maxwell waves with the gravitational field. Nevertheless, their results do neither predict the EM wave polarization--black hole rotation coupling nor the superluminal propagation of electromagnetic waves.

The propagation of light along null geodesics is an exact result for plane waves propagating in vacuum flat spacetimes \cite{hojmanIso}, or an approximate geometrical optics limit for light propagation in curved spacetimes.
However, we have shown that beyond this approximation, in G\"odel and Kerr spacetimes, this behavior changes and the electromagnetic  wave does not follow  null geodesics.
This is due to the rotational nature of the spacetime (i.e., non-diagonal components of the metric) that produce an effective anisotropic medium where the photons propagate.  In those cases, the nature of the EM field is non-null in general for the approximated solutions presented in previous sections. However, a more refined calculation for the exact  invariants for the EM waves is lacking here, and it is left for future works.

Therefore, there appears to be {\textit{no single speed of light}}, but electromagnetic radiation propagates with different speeds (different $K_\mu K^\mu$ values) depending on the interaction of its polarization with the gravitational background. One may still define the speed of light which corresponds to waves which  propagate obeying  $K_\mu K^\mu= 0$.
This remarkable result could have strong implications in  astrophysics, where accurate measurements of the speed of light  are crucial.

Yet, the more striking new idea emerges from Eq.~\eqref{DispK2} for Kerr metric, where the importance of the light polarization in the wave propagation is explicitly shown. This new phenomenon also  exhibits the spacelike character of the wave vectors. It is well known that the polarization of a wave affects its propagation properties in a medium \cite{born,chen}, but to the best of our knowledge, no such predictions have been reported for EM waves travelling in vacuum in curved spacetime. Nevertheless, the relevance of the polarization of an electromagnetic wave in flat spacetime has been put in manifest in experiments showing that structured  light waves (not plane waves)  can travel in vacuum slower than speed of light \cite{Giovannini,bareza}. Furthermore, it has been recently observed that quantum effects produce birefringence of EM waves in vacuum \cite{mignani}.

Images of celestial objects in which black holes act as gravitational lenses might be obtained in the near future by using interferometers such as the GRAVITY ESO VLTI described in Refs.~\cite{lens1,lens2}, for instance.  One way to put our prediction to test is to get images of celestial objects which
 emit both circular polarizations, so that their different propagation can be directly compared by their different states of polarization. At least two different images should be obtained as a consequence of the different trajectories followed by waves with different kinds of polarization.
Long wavelengths, i.e., radio frequencies waves seem to be the best candidates
to show such an effect.

In conclusion, the gravitational field can alter the  path that an electromagnetic  wave follows.  For the above discussed phenomena to be significant, in general the  EM wavelength and the curvature length scale must be comparable for the non-geodesic propagation of light. One can imagine that the measurements of these effects on astrophysical length scales would be difficult due to the very low frequencies of EM waves. However, due to the effective medium produced by gravity, the wave is dispersive, i.e. the relation between wavelengths and frequency is not trivial, and the frequency can be higher than expected. On the other hand, there are other effects not directly related to the frequency but to the EM wave polarization, such as the coupling of polarization to the black hole rotation. And there even exists a coupling between polarization and the anisotropy of spacetime itself \cite{bianchiasenjohojman}.
Thus, all the above is extremely  relevant for the understanding of the evolution and interaction of electromagnetic fields with matter at large scales in the Universe.

\begin{acknowledgments}
The authors wish to express their gratitude to Andr\'es Gomberoff and Benjamin Koch for fruitful discussions. They also gratefully acknowledge the referees' comments and suggestions which helped to improve this article significantly. F.A.A. thanks CONICyT-Chile for Funding No. 79130002.
\end{acknowledgments}

\end{document}